\documentclass[preprint]{revtex4-2}
\usepackage{color}
\usepackage{graphicx}
\usepackage{amssymb}
\usepackage{amsmath}
\usepackage{hyperref}

\begin{document}
	 
\title{Generating ultra compact boson stars with modified scalar potentials}

\author{Sarah Louisa Pitz}
\email{pitz@itp.uni-frankfurt.de}
\affiliation{Institut f\"ur Theoretische Physik, Goethe Universit\"at, Max-von-Laue Straße 1, D-60438 Frankfurt am Main, Germany}

\author{J\"urgen Schaffner-Bielich}
\email{schaffner@astro.uni-frankfurt.de}
\affiliation{Institut f\"ur Theoretische Physik, Goethe Universit\"at, Max-von-Laue Straße 1, D-60438 Frankfurt am Main, Germany}

\date{\today}

\begin{abstract}
The properties of self-interacting boson stars with different scalar potentials going beyond the commonly used $\phi^4$ ansatz are studied.
The scalar potential is extended to different values of the exponent $n$ of the form $V \propto \phi^n$. 
Two stability mechanism for boson stars are introduced, the first being a mass term and the second one a vacuum term.
We present analytic scale-invariant expressions for these two classes of EOS.
The resulting properties of the boson star configurations differ considerably from previous calculations.
We find three different categories of mass-radius relation: the first category resembles the mass-radius curve of self-bound stars, 
the second one those of neutron stars and the third one is the well known constant radius case from the standard $\phi^4$ potential.
We demonstrate that the maximal compactness can reach extremely high values going to the limit of causality $C_\text{max} = 0.354$ 
asymptotically for $n\to\infty$. The maximal compactnesses exceed
previously calculated values of $C_\text{max}=0.16$ for the standard $\phi^4$-theory and $C_\text{max}=0.21$ for vector-like interactions
and is in line with previous results for solitonic boson stars.
Hence, boson stars even described by a simple modified scalar potential in the form of $V \propto \phi^n$ 
can be ultra compact black hole mimickers where the photon ring is located outside the radius of the star.
\end{abstract}

\maketitle
\newpage

\section{Introduction}\label{incanus}

Dark matter plays a crucial role in explaining certain phenomena in cosmology and astrophysics on large scales, 
where self-interacting dark matter provides an explanation for small-scale structure observations such as the core-cusp and the missing satellites problem \cite{Tulin:2017ara}. 
By including self-interactions dark matter can form compact objects, such as boson stars. Boson stars are self-gravitating spheres, described by complex scalar fields, see \cite{Liebling:2012fv} for a review. 
These kind of compact objects were discussed for the first time by Wheeler in 1955 \cite{Wheeler:1955zz} with self-gravitating, non-interacting spheres made of bosons, which he named geons. 
However, these configurations turned out to be unstable. In order to obtain stable stars one needs time dependent solutions of the Klein-Gordon equation \cite{Derrick:1964ww}. 
These stable boson stars were usually of microscopic sizes for the non-interacting case \cite{Kaup:1968zz,Ruffini:1969qy}. 
Boson stars of astrophysical sizes were found 
for solitonic boson stars \cite{Lee:1986tr,Lee:1986ts,Friedberg:1986tp,Friedberg:1986tq},
by introducing a self-interaction potential of the form $V = \lambda\phi^4$ \cite{Colpi:1986ye}, and
for repulsive vector-like self-interactions proportional to the density squared \cite{Agnihotri:2008zf}.
Generic self-interactions have been considered in \cite{Schunck:1999zu}.
The properties of boson stars with self-interactions have been investigated
in detail using a $\phi^4$-potential or vector-like self-interactions \cite{Eby:2015hsq,Kouvaris:2015rea,Maselli:2017vfi,Cassing:2022tnn}.
Boson stars can be built via standard structure formation from the early universe 
\cite{Khlopov:1985jw,Chang:2018bgx}. So far the LIGO-Virgo collaboration has only detected one event of a neutron star-neutron star merger, GW170817 \cite{LIGOScientific:2017vwq}, confirmed by a gamma-ray burst and the optical afterglow. 
A significant amount of the gravitational wave sources measured by the LIGO-Virgo collaboration is located in the mass range between the lightest black hole of $5M_\odot$ and the heaviest neutron star \cite{LIGOScientific:2021djp}. 
These compact objects could possibly be exotic stars since the most massive neutron stars measured so far have masses of $2.01 \pm 0.04 M_\odot$ respectively $2.08 \pm 0.07 M_\odot$, constrained by observations of the radio pulsars PSR J0348+0432 and PSR J0740+6620 \cite{Antoniadis:2013pzd,Fonseca:2021wxt,Riley:2021pdl} as well as $2.35 \pm 0.17 M_\odot$, constrained from optical observations of the black widow pulsar PSR J0952-0607 \cite{Romani:2022jhd}.
For example, in the gravitational event GW190814 one compact object has a mass of $2.6 M_\odot$, exceeding all known neutron star masses considerably \cite{LIGOScientific:2020zkf}. 
Boson stars can be observed by the emission of gravitational waves from
a merger of two boson stars or one boson star with a neutron star.
Moreover gravitational waves from boson star-boson star merger can be distinguished from other mergers by their echo in gravitational waves \cite{Cardoso:2016oxy,Maselli:2017tfq,Mark:2017dnq,Sennett:2017etc,Palenzuela:2017kcg,Bezares:2017mzk,Toubiana:2020lzd,Bezares:2022obu}.
Also collisions of neutron stars with boson stars can send out gravitational waves that are detectable by future telescopes \cite{Dietrich:2018bvi}. Here an unusual deviation by e.g. an extremely small value of the tidal deformability $\Lambda$ can indicate such an exotic encounter.
We will find below that the maximum compactness $C=GM/R$ for boson stars is $C= 0.354$. Since the tidal deformability goes like $\Lambda \sim k_2 C^{-5}$ with a Love number of $k_2\approx 0.03$ or smaller
the value for boson stars can be as low as of the order of $1$, which is much smaller compared to the standard values 
for ordinary neutron stars. For comparison we note that the limit $\Lambda < 720$ has been extracted from the neutron star merger event GW170817 \cite{LIGOScientific:2018hze}. 
Another possibility of gravitational wave sources are ordinary neutron stars containing a bosonic dark matter fluid \cite{Bauswein:2020kor,Lee:2021yyn,Karkevandi:2021ygv,Giangrandi:2022wht,Rutherford:2022xeb,Shakeri:2022dwg,Diedrichs:2023trk}. 
These two-fluid compact objects could contain dark matter cores which are influencing the macroscopic properties of the star. 
With the advent of observing supermassive black holes by radio observations of the Event Horizon Telescope (EHT) collaboration one can study the accretion disk of these objects
in detail \cite{EventHorizonTelescope:2019dse,EventHorizonTelescope:2022wkp}. 
The precision of these measurements allows to distinguish between black holes and other compact objects, as e.g. boson stars. 
In comparison to black holes boson stars don't have a shadow since light does not interact with dark matter and thus passes through the star.
This feature has been used in the analysis of the first pictures from black hole M87* and SGR A* by radio observations to constrain models
of boson stars as black hole mimickers \cite{Olivares:2018abq}. In our work we are extending the standard $\phi^4$ self-interaction potential of bosons 
to a general one of the form $\phi^n$ with arbitrary values of the power $n$. Examples like the three gluon interaction for $n = 3$ and the Wess-Zumino-Witten term in the chiral Lagrangian to describe the coupling between two kaons and three pions for $n = 5$, motivate the study of other values of $n$.
By choosing this generalized self-interaction potential we find that compact stars with massive bosons can have different forms of their mass-radius relations, with some of them being similar to those of neutron stars or self-bound stars. We find curves that are not constrained to the constant radius case for the $\phi^4$-potential and the one of vector-like self-interactions. 
We will also show that their compactness can exceed those of neutron stars and goes asymptotically to the limit of causality with a compactness of $C = 0.354$ for large values of $n$. 
These extreme values of compactness are similar to the ones found in a recent work on solitonic boson stars \cite{Boskovic:2021nfs}
and lead to new opportunities in the search of self-interacting boson stars via merging boson stars
and thus in the search of self-interacting dark matter. 
For a first investigation of ultra compact solitonic boson star mergers we refer to \cite{Bezares:2022obu}. 
The outline of the paper is as follows: first the theoretical basis from classical field theory for complex scalar fields and general relativity is summarized. Then equations of state (EOS) with different stability mechanisms for boson stars will be introduced. We discuss two stabilizing mechanisms: a standard mass term in the Lagrangian and a vacuum energy in the potential without a mass term. 
Finally the mass-radius curves as well as the compactness are presented.

%%%%%%%%%%%%%%%%%%%%%%%%%%%%%%%%%%%%%%%%%%%%%%%%%%%%%%%%%%%%%%%%%%%%%%%

\section{Theoretical Framework}

\subsection{Scale-invariant equation of state from a classical scalar potential}  
 
Assuming a complex scalar field for the description of a bosonic matter, a suitable Lagrangian reads 
as follows:
\begin{equation} \label{lagrangian}
    \mathcal{L} = \partial_\mu \phi^*\partial^\mu \phi +m^2\phi^*\phi- V	
\end{equation}
 where $V$ represents the potential, $m$ the mass of the boson and $\phi$ the complex scalar field. 
  The equation of motion is then given by
 \begin{equation}\label{eom}
	(\partial_\mu \partial^\mu + m^2)\phi = -\frac{\partial V}{\partial \phi^*}.
\end{equation}
For the investigation of the potential we are going to present analytic general EOS which only depend on the exponent $n$. 
In this work we assume an ideal fluid for the bosons and calculate the EOS adopting a flat space-time. 
 Assuming flat space-time is justifiable in a local density approach. Without any interaction potential the radius of curvature of the boson star 
 is of the same order as the Compton wavelength of the massive boson ($r \propto M_P/m_b^2$ with the Planck mass $M_P$ and the mass of the boson $m_b$) \cite{Colpi:1986ye}, which means space-time is strongly curved. The opposite is given with a strong interaction potential with an interaction strength $\lambda$. The radius of curvature increases with $M_P/m_b \sqrt{\lambda}$ \cite{Colpi:1986ye}, which allows to consider flat space-time. Moreover the scalar field only varies on a large scale, so the gradient of the field can be neglected. This is given, when $M_P/m_b\sqrt{\lambda} \gg 1$ \cite{Colpi:1986ye}.
 Starting off with the energy-momentum tensor to calculate the equation of state 
 \begin{equation}
		T_{\mu \nu} = -\eta_{\mu \nu} {\cal L} + \sum_{\phi,\phi^*}\frac{\partial \mathcal{L}}{\partial(\partial^\mu\phi_i)}\partial_\nu \phi_i.
\end{equation}
Together with equation (\ref{lagrangian}) we get
\begin{equation} \label{Tmn_L}
    T_{\mu \nu} = - \eta_{\mu \nu} {\cal L} + \partial_\mu \phi^* \partial_\nu \phi + \partial_\mu \phi \partial_\nu \phi^* 
    + \frac{\partial V}{\partial(\partial^\mu \phi)}\partial_\nu \phi
    + \frac{\partial V}{\partial(\partial^\mu \phi^*)}\partial_\nu \phi^*.    
\end{equation}
The last two terms in equation (\ref{Tmn_L}) vanish since the potentials used in this work do not depend on the derivatives of the scalar field. 
The energy-momentum tensor reduces to the form for an ideal fluid:
	\begin{eqnarray}
		T^\mu_\nu = \begin{pmatrix}
			\varepsilon & 0 & 0 & 0 \\
			0 & p & 0 & 0\\
			0 & 0 & p & 0\\
			0 & 0 & 0 & p\\
		\end{pmatrix}
	\end{eqnarray}
 with the energy density $\varepsilon$ and the pressure $p$. 
Calculating $T_{00}$ and $T_{ii}$ and making use of the ansatz $\phi = \phi_0 e^{-i\omega t}$ leads to the following equations:
	\begin{eqnarray}
		T_{00} &=& \phi_0^2\omega^2 + m^2\phi_0^2 + V \label{T_00}  \\ 
		T_{ii} &=& \phi_0^2 \omega^2 - m^2\phi_0^2 - V \label{T_ii}
	\end{eqnarray}
 where $\omega$ and $t$ denote energy and time respectively. Using the ansatz together with equation (\ref{eom}) gives a relation between $\omega$ and $m$:
 \begin{equation}
     \omega^2 \phi_0^2 = \phi_0^2 m^2 + \frac{\partial V}{\partial \phi^*} \phi^*.
 \end{equation}
 Making use of this relation leads to the desired expressions for $\varepsilon$ and $p$:
	\begin{eqnarray}
		T_{00} &=& \varepsilon = 2m^2 \phi_0^2 + \frac{\partial V}{\partial \phi^*}\phi^* + V \label{general form mass}\\
		T_{ii} &=& p = \frac{\partial V}{\partial \phi^*}\phi^* - V. \label{general form mass 2}
	\end{eqnarray}
To obtain dimensionless quantities we divide equations (\ref{general form mass}) and (\ref{general form mass 2}) 
by a factor $1/m^4$
\begin{eqnarray}
    \varepsilon' &=& 2\phi_0^{'2} + \frac{\partial V'}{\partial \phi^*}\phi^* + V'\\
    p &=& \frac{\partial V'}{\partial \phi^*}\phi^* - V'
\end{eqnarray}
with $\varepsilon' = \varepsilon / m^4$, $V' = V / m^4$ and $p' = p / m^4$.

\subsection{Tolman-Oppenheimer-Volkoff (TOV) equations}\label{TOVsec}

By solving the TOV equations one obtains the mass and the radius of an compact object, which is necessary in order to calculate the mass-radius curves: 
	\begin{eqnarray}
		\frac{dp}{dr} &=& - G\frac{m_r(r)\varepsilon(r)}{r^2}\biggl( 1 + 		
  \frac{p(r)}{\varepsilon(r)} \biggr)\biggl( 1 + \frac{4\pi r^3p(r)}{m_r(r)}\biggr) \biggl( 1 - \frac{2Gm_r(r)}{r} \biggr)^{-1} \\
		\frac{dm_r(r)}{dr} &=& 4 \pi r^2 \varepsilon(r)
	\end{eqnarray}
 with the pressure $p(r)$, radius $r$, the gravitational constant $G$, the mass $m_r$ inside a sphere of radius $r$ and the energy density $\varepsilon(r)$. Since all the calculations are dimensionless, the TOV equations need to be rescaled. Applying the following scaling relations for the mass and the radius
	\begin{eqnarray}
		m_r &=& (G^3\cdot \varepsilon_0)^{-1/2}m_r'\\
		r &=& (G\cdot \varepsilon_0)^{-1/2}r'
	\end{eqnarray}
 leads to the following form:
	\begin{eqnarray} \label{TOV p}
		\frac{dp'}{dr'} &=& -\frac{m_r'\varepsilon'}{r'^2}\biggl(1 + 		\frac{p'}{\varepsilon'}\biggr)\biggl(1 + \frac{4\pi r'^3 p'}{m_r'}\biggr)\biggl(1 - 		\frac{2m_r'}{r'}\biggr)^{-1} \\
		\frac{dm_r'}{dr'} &=& 4 \pi r'^2 \varepsilon'.
	\end{eqnarray}
 with $\varepsilon_0$ being a constant with dimension of an energy density. Also the pressure and the energy density need to be rescaled:
	\begin{eqnarray}
		p' &=& \frac{p}{\varepsilon_0}\\
		\varepsilon' &=& \frac{\varepsilon}{\varepsilon_0}.
	\end{eqnarray} 
 Please note that in natural units the energy density and the pressure have the same dimension. 
 For e.g.\ the case of non-interacting massive bosons with a mass $m$ one would choose $\varepsilon_0=m^4$.

\subsection{Dimensionless Equation of State}\label{StabMechan}

Since all calculations are independent on units, we need to derive an equation of state that satisfies this requirement. 
We use in the following two different kind of generalized scalar potentials: one with a mass term and one with a vacuum term but without a mass term. It turns out that these two kinds of potentials lead to stable compact star configurations.
We note in passing that without a mass and a vacuum term the compact star configurations are unstable.
For the two different stability mechanisms the EOS differ considerably as discussed below.
The starting point for the following calculations is equation (\ref{general form mass}), respectively equation (\ref{general form mass 2}). 
For the scalar potential of the form
\begin{equation}
V = \frac{\lambda}{2^{n/2}} \left(\phi^* \phi\right)^{n/2}
\end{equation}
we get
 \begin{eqnarray}
     p' &=& \frac{\lambda'}{2^{n/2}} (\phi_0')^n \left(\frac{n}{2} - 1 \right)\\
     \varepsilon' &=& 2\phi_0'^2 + \frac{\lambda'}{2^{n/2}} (\phi_0')^n \left(\frac{n}{2} + 1 \right)
 \end{eqnarray}
with the dimensionless coupling strength $\lambda' = \lambda/m^{4-n}$ 
and the dimensionless scalar field $\phi_0' = \phi_0/m$. 
In order to obtain an analytic expression for the EOS, we can express the pressure in terms of the scalar field and insert it
into the expression for the energy density. The EOS can be further simplified by rescaling the pressure and the energy density with a factor $\lambda'(n/2-1)$ to the form
	\begin{equation}\label{EoS phi^n mass}
		\varepsilon' = p'^{2/n} + \frac{n+2}{n-2}\,p'.
	\end{equation}  
 It is evident that the EOS (\ref{EoS phi^n mass}) is restricted to $n>2$. 

 %\subsubsection{Equation of State for the Vacuum Constant Case}

For the second form of the EOS studied we introduce a vacuum term $V_0$.
We start with equation (\ref{general form mass}) and (\ref{general form mass 2}) but setting the mass term to zero. 
The potential in this case is given by 
\begin{equation}
V = \frac{\lambda}{2^{n/2}} \left(\phi^* \phi\right)^{n/2} + V_0.
\end{equation}
We obtain for the pressure and energy density  
 \begin{eqnarray}
     \varepsilon &=& \frac{\lambda}{2^{n/2}}\phi_0^n\left(\frac{n}{2} + 1 \right) + V_0\\
     p &=& \frac{\lambda}{2^{n/2}}\phi_0^n \left(\frac{n}{2} - 1 \right) - V_0.
 \end{eqnarray}
 Combining these two equations gives an EOS which is independent 
 on the interaction strength $\lambda$
 	\begin{equation}\label{eos vac term1}
		\varepsilon = \frac{n+2}{n-2}\,p + \frac{2n}{n-2}\,V_0.
	\end{equation}
Interestingly, this EOS is the one for self-bound stars of the form
\begin{equation}
    p = c_s^2 \left( \varepsilon - \varepsilon_\text{vac} \right)
\end{equation}
where the pressure vanishes at a nonvanishing vacuum energy density $\varepsilon_\text{vac}$.
The prefactor $c_s^2$ stands for the speed of sound. 
As one can see from eq.~(\ref{eos vac term1}) different values of $c_s^2$,
i.e.\ different stiffnesses of the EOS,
emerge from the chosen value of the power $n$ in the scalar potential.
Note that the self-bound EOS for has been derived from interacting bosonic matter. 
One arrives at the MIT bag EOS by setting $n=4$ in eq.~(\ref{eos vac term1}) so that $c_s^2=1/3$.
These similar EOS are based on a completely different descriptions of the matter, demonstrating that the EOS is composition blind so that the results from general relativity
do not depend on the underlying microphysics, as dictated from the strong equivalence principle.
Rescaling the EOS with  $\varepsilon' = \varepsilon/V_0$ and 
$p' = p/V_0$ results in a dimensionless EOS. 
A further rescaling with the factor $2n/(n-2)$ gives the final dimensionless EOS of the form
  	\begin{eqnarray}\label{eos vac term DR}
		\varepsilon' = \frac{n+2}{n-2}\,p' + 1.
	\end{eqnarray}
This EOS is quite similar to the one with the mass term (see equation (\ref{EoS phi^n mass})). 
The part linear in $p$ stays the same but the second part differs and is now a constant.

%%%%%%%%%%%%%%%%%%%%%%%%%%%%%%%%%%%%%%%%%%%%%%%%%%%%%%%%%%%%%%%%%%%%%%%

\section{Results: Mass-Radius Curves and Compactness}\label{MRsec}

By solving the TOV equations together with the derived EOS, we obtain the corresponding mass-radius curves.

\begin{figure}
	\centering
	\includegraphics[width=0.65\linewidth]{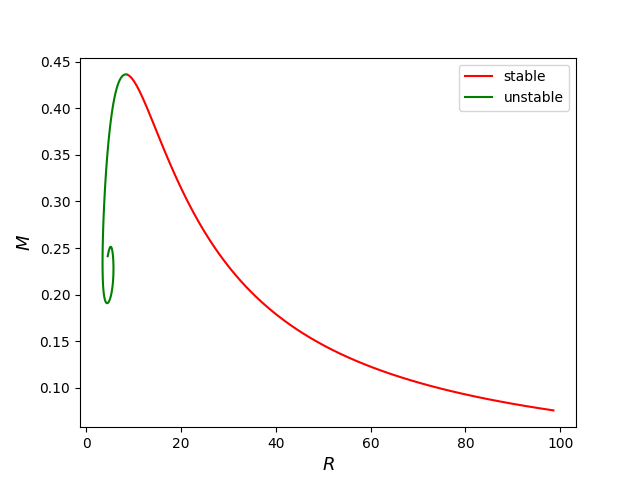}
	\caption{Mass-radius curve for $V \propto \phi^3$ with a mass term in the Lagrangian in dimensionless units. 
 The mass decreases with increasing radius in the stable branch as $M \propto R^{-1}$.}
	\label{fig:phi3neuneu}
\end{figure}

\begin{figure}
    \centering
    \includegraphics[width=0.65\linewidth]{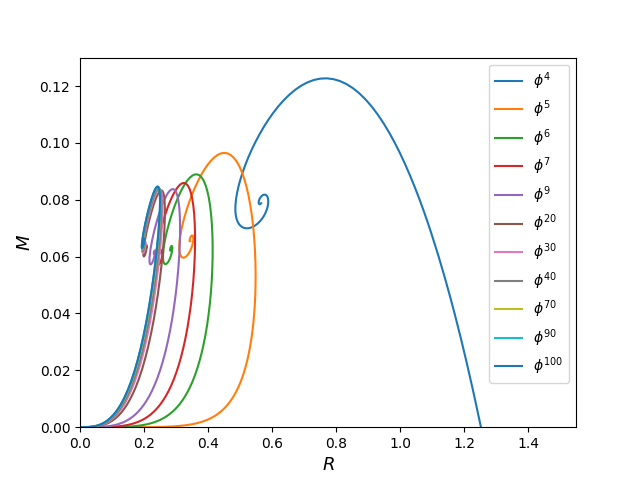}
    \caption{Mass-radius curves for $V \propto \phi^n$ for $n=4$ up to $n=100$ with a mass term in the scalar potential, shown in dimensionless units.
     The maximum mass and radius decrease with $n$. The mass-radius curve for $n=4$, i.e.\ for the standard $\phi^4$-potential, 
     has a different shape compared to the curves with $n>4$ since the radius goes to a constant value when the mass goes to zero. 
     The other mass-radius curves resemble those of self-bound stars as they start at the origin.}
    \label{fig:kombiplot4_1000neuneu}
\end{figure}

Figures \ref{fig:phi3neuneu} and \ref{fig:kombiplot4_1000neuneu} show the mass-radius curves for the case with a mass term
for different values of the power $n$. The case $n=3$ is plotted in a separate figure, see Fig.~\ref{fig:phi3neuneu}, 
due to a different magnitude of the maximum mass and the corresponding radius. 
Additionally this curve differs in its shape compared to curves with larger values of $n$, since the mass decreases with increasing radius,
in a fashion known for e.g.\ neutron stars.
Nevertheless its compactness is lower than the ones for larger values of the power $n$. 
Note that the solutions to the left of the maximum are unstable as depicted in Fig.~\ref{fig:phi3neuneu}. 
The mass-radius curves for $n=4$ starts for vanishing mass at a nonvanishing radius, as it is typical for a mass-radius relation
with a radius independent on the mass. On the other hand, the mass-radius curves for the cases with $n=5$ and larger start at the origin, 
i.e.\ at vanishing mass and radius. The shape of these mass-radius curves look like the ones of self-bound stars where the mass increases
with $R^3$. However, the underlying EOS does not exhibit a nonvanishing energy density at vanishing pressure.

In summary three types of solutions are identifiable, for $n = 3$, $n = 4$ and $n > 4$. 
The curves differ in their behaviour in the limit of small masses. 
The curve with $n = 3$ goes to infinite radius, 
the curve with $n = 4$ goes to a constant value and the curves with $n > 4$ are going to zero for small masses.
In order to understand their behaviour it is useful to consider the limit of small pressure $p$ for the EOS. 
Equation (\ref{EoS phi^n mass}) then simplifies to:
\begin{equation}\label{mR rel eos}  
	\varepsilon \approx p^{2/n}.
\end{equation}
 These different shapes can be understood by having a look at the mass-radius relation of a sphere in hydrostatic equilibrium
 for a polytropic EOS of the form $p \propto \rho^\Gamma$, with $\rho$ being the mass density and $\Gamma$ a constant.
 In the nonrelativistic limit at low density, the energy density is simply given by the mass density 
$\varepsilon = \rho$.
The mass $M$ and the radius $R$ for a polytrope are related by (see e.g.\ \cite{Schaffner-Bielich:2020psc}):
\begin{equation}\label{MRrelationPoly}
		M^{2-\Gamma}\cdot R^{3\Gamma - 4} \propto \text{const.}
\end{equation} 
Equation (\ref{mR rel eos}) describes a polytropic EOS and thus gives the possibility to make use of equation (\ref{MRrelationPoly})
by setting $\Gamma = n/2$.

\begin{table}\label{tableMRrelation}
	\centering
	\renewcommand{\arraystretch}{1.2}
	\begin{tabular}{|c|c|c|c|c|c|c|c|}
		\hline
		$\boldsymbol{n}$ & $\boldsymbol{3}$ & $\boldsymbol{4}$ & $\boldsymbol{5}$ & $\boldsymbol{6}$ & $\boldsymbol{7}$ & $\boldsymbol{8}$ & $\boldsymbol{\infty}$\\[.5ex]
		\hline
		$\boldsymbol{\Gamma}$ & $\frac{3}{2}$ & $2$ & $\frac{5}{2}$ & $3$ & $\frac{7}{2}$ & $4$ & $\infty$\\[.5ex]
		\hline
		$\boldsymbol{M}$-$\boldsymbol{R}$ \textbf{relation} & $M \propto R^{-1}$ & $R \propto \text{const.}$ & $M \propto R^{7}$ & $M \propto R^{5}$ & $M \propto R^{13/3}$ & $M \propto R^4$ & $M\propto R^3$\\[.5ex]
		\hline 
	\end{tabular}
	\caption{Mass-radius relations for different values of $n$ derived by using equation (\ref{MRrelationPoly}).}	
\end{table}

 In table \ref{tableMRrelation} we calculated the mass-radius relations for different values of $n$. 
 We can reproduce that the mass decreases in the limit of large radii in the case of $n = 3$ as $M\propto{}R^{-1}$.
 Furthermore table \ref{tableMRrelation} confirms a constant value of the radius $R$ for small pressures, respectively small masses, for the case $n=4$. 
 The remaining cases (with $n > 4$) are described by slightly different mass-radius relations. 
 However, the mass vanishes in all these cases in the limit of $R \rightarrow 0$, making them look like the mass-radius curves of self-bound stars. In the limit of $n\to\infty$ one recovers the mass-radius relation of
 an incompressible fluid, i.e.\ $M/R^3 = \text{const.}$ as for self-bound stars. 
 Please note that these relations are valid for small pressure $p$ which corresponds to small masses. For higher pressures $p$ the first term in equation (\ref{EoS phi^n mass}) dominates so that
 the mass-radius relations shown in the table \ref{tableMRrelation} do not hold anymore. As seen in 
 Figs.~\ref{fig:phi3neuneu} and \ref{fig:kombiplot4_1000neuneu}, 
 this is the case for the configurations close to the maximum mass.
 
 Another way to understand the shape of the curves is by having a look at their slope. 
 By rearranging equation (\ref{MRrelationPoly}) we obtain the relation
\begin{eqnarray}
	\frac{\text{d}\log M}{\text{d}\log R} =  \frac{3\Gamma - 4}{\Gamma-2} \label{linear func}.
\end{eqnarray}
We can now identify the right-hand side of the equation with the slope of the curve $m$. 
For $n=3$ one finds $m = -1$, for $n=4$ the slope goes to infinity ($m \rightarrow \infty$) 
and the case $n>4$ gives constant positive values for $m$. 
Again we can confirm the shape of the curves with our numerical results.
As already mentioned before the curves for $n>4$ look similar to those of self-bound stars, 
where gravity is not needed to ensure stability. 
self-bound stars are characterized by a nonvanishing value of the energy density at zero pressure. 
The shape of the mass-radius relation is determined by a constant energy density, ensuring that $M \propto R^{3}$ holds. 
This explains why the mass-radius curves of those stars are located at the origin. 
However, we stress that the stars here at not purely self-bound, since they need gravity to remain stable. 

\begin{figure}
	\centering
	\includegraphics[width=0.65\linewidth]{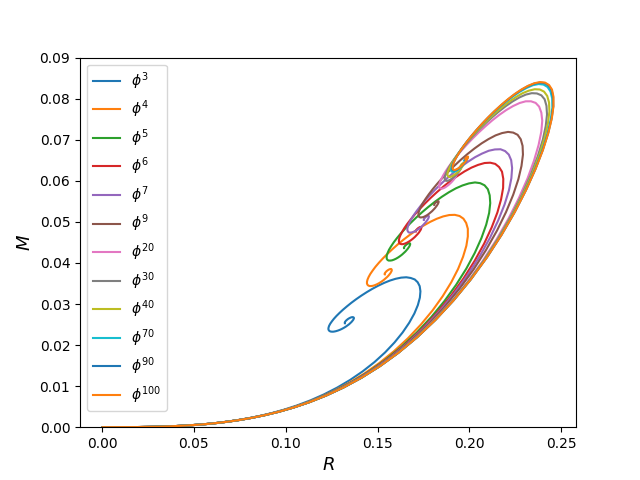}
	\caption{Mass-radius curves for $V \propto \phi^n$ stabilized with a vacuum constant in dimensionless units. 
 All curves have a similar shape and the same behavior in the limit of small masses, i.e.\ the radius vanishes for small masses. The maximum mass and radius increase with $n$.}
	\label{fig:MRcurvesDR}
\end{figure}

The mass-radius curves for the scalar potential without a mass term but with a vacuum term are depicted
in Fig.~\ref{fig:MRcurvesDR}.
Contrary to the mass term case, these mass-radius curves only lead to one type of solution. 
The mass vanishes for small radii and increases with increasing radius. The mass-radius curves for the different values of $n$ are lying on top of each other for small radii. This feature originates from the dominant behaviour of the vacuum term in the EOS at low densities, which is independent on the value of $n$.
The energy density stays nearly constant as it is just given by the vacuum energy density 
resulting in a mass-radius curve of the form $M/R^3 = \text{const.}$, i.e.\ the familiar one for self-bound stars.
The maximum masses and radii increase with higher values of $n$.
For both types of EOS we also investigated
their dependence on the speed of sound squared $c_s^2$, which is defined as the derivative of the pressure with respect to the energy density:
   	\begin{eqnarray}
 		c_s^2 = \frac{\partial p}{\partial \varepsilon}.
	 \end{eqnarray}  
This gives the following simple relation between the equation of state and $c_s^2$ for the vacuum term case:
 	\begin{eqnarray}
 		c_s^2 &=& \frac{n-2}{n+2}
 	\end{eqnarray} 
while for the case with a mass term one arrives at the relation:
    \begin{eqnarray}
        c_s^2 = \left(\frac{2}{n}\,p^{\frac{2-n}{n}} + \frac{n+2}{n-2}\right)^{-1}.
    \end{eqnarray}
One realizes that the EOS with a vacuum term results in a constant $c_s^2$. 
The EOS with a mass term has an increasing $c_s^2$ with increasing pressure 
(strictly speaking for $n>2$ which is the case in our studies). In the limit of $p\to\infty$
one recovers the speed of sound squared of the case with a vacuum term. Hence, for a given power $n$,
$c_s^2$ will always be larger for the EOS with a vacuum term compared to the one with a mass term.
Thereby, the EOS with a vacuum term will be stiffer, for a fixed value of $n$, compared to the one with a mass term.
We also see, that $c_s^2$ increases monotonically with the power $n$ reaching $c_s^2=1$ in the limit $n\to\infty$.
 
\begin{figure}
	\centering
        \includegraphics[width=0.65\linewidth]{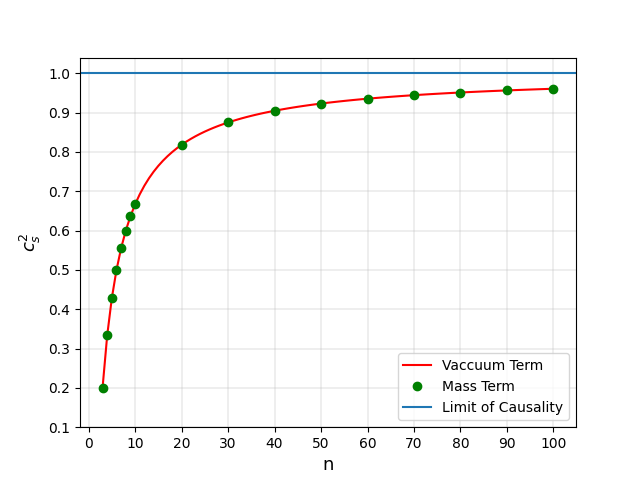}
	\caption{The speed of sound squared $c_s^2$ plotted against the exponent of the potential $n$. $c_s^2$ increases with increasing $n$ and goes asymptotically to $c_s^2=1$. This is the stiffest possible equation of state, where the speed of sound is equal to the speed of light.}
	\label{fig:cs2vsnfit}
\end{figure}
In figure \ref{fig:cs2vsnfit} we plot the results for the maximal value of $c_s^2$ in the center of the maximum mass configuration, 
compared for both cases of the EOS. We find that the values of $c_s^2$ are in good agreement 
but not identical as expected. This feature indicates that the maximum mass configuration for the EOS with a mass term
has a $c_s^2$ close to their asymptotic limit for large pressures. For large values of $n$ the speed of sound squared 
reaches the causal limit $c_s^2=1$ where the speed of sound squared equals the speed of light.

In addition to $c_s^2$ we also studied the compactness, respectively the maximum compactness $C_\text{max}$, of the the resulting compact star configurations, defined as:
	\begin{eqnarray}
		C = \frac{M}{R}, 
	\end{eqnarray}
 with $M$ and $R$ representing the dimensionless quantities calculated by solving the TOV equations. 
 In the following we demonstrate that $C_\text{max}$ can exceed the maximum value for quark stars ($C_\text{max} = 0.271$), as well as the compactness needed to place the photon orbit outside the star ($C_\text{max} = 1/3$) 
 and that the maximum compactness goes asymptotically to the limit of causality $C_\text{max} = 0.354$ for large
 values of the power $n$. The existence of a photon sphere for boson stars increases the probability of misinterpretating them as black holes, since black holes always have a photon sphere around them as a tell-tale signature \cite{Claudel:2000yi}. 

\begin{figure}
	\centering
	\includegraphics[width=0.65\linewidth]{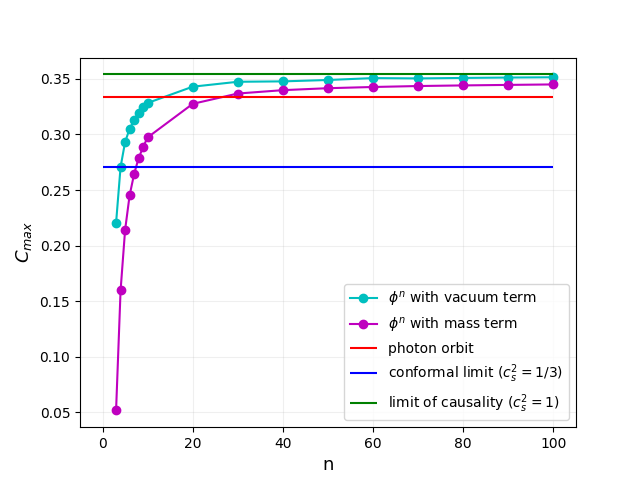}
	\caption{The maximum compactness for the $V \propto \phi^n$ potential with a mass and a vacuum term plotted against $n$. 
 Furthermore different limits for compact objects are included in the plot. 
 The red line represents the limit for the photon sphere being outside the radius of the star ($C = 1/3$).
 The green line displays the maximum compactness for causal EOS ($C = 0.354$) (causal in the sense that the speed of sound cannot exceed the speed of light). 
 The blue line represents the maximum compactness for self-bound stars with a conformal EOS where $c_s^2=1/3$ 
 ($C = 0.271$).}
    \label{fig:cmaxvsnneuebeschriftung}
\end{figure}
The results for both cases of the EOS are depicted in figure \ref{fig:cmaxvsnneuebeschriftung}. 
We find that for higher values of $n$ the maximum compactness increases monotonically.
The steepest slope of the curves is in the range of $n = 3$ up to $n = 10$.  From that point on the maximum compactness
goes asymptotically to an upper limit. This boundary is given by the limit of causality $c_s^2 = 1$ with
$C = 0.354$, where the speed of sound is equal to the speed of light. 

This feature implies that the compactness can be greater than $C = 1/3$.
The photon sphere for the Schwarzschild metric lies at $R = 3M$ 
with a corresponding minimal compactness of $C = M/R = 1/3$.
According to Fig.~\ref{fig:cmaxvsnneuebeschriftung} one realizes that the photon orbit 
can lie outside of the boson star for sufficiently large values of $n$, 
producing a light ring.
A similar feature has been seen for solitonic boson stars where the maximum compactness
also reaches asymptotically the one for the causal limit \cite{Boskovic:2021nfs}. 
We conclude that solitonic boson stars as well as self-interacting boson stars studied 
here can constitute black hole mimickers 
with a maximal compactness in excess of $C=1/3$.

Figure \ref{fig:cmaxvsnneuebeschriftung} shows that boson stars can reach higher $C_\text{max}$ than expected. 
These values exceed those of standard neutron stars 
(without a phase transition typically $C_\text{max} = 0.2$ to 0.3)
or other exotic compact objects like quark stars where $C=0.271$, 
(see e.g.\ the discussions in 
\cite{Lattimer:2012nd,Schaffner-Bielich:2020psc}).
We find that the maximal compactness for $n = 4$ in the vacuum term case is $C = 0.271$ which is in 
agreement with the values for quark stars quoted above.
This was expected since equation (\ref{eos vac term1}) for this value of $n$ is
$\varepsilon = 3p + \text{const.}$
which is the equation of state for an ultrarelativistic ideal gas of quarks with nonvanishing vacuum energy,
given by the MIT bag constant, or more general a conformal EOS with $c_s^2=1/3$.

%%%%%%%%%%%%%%%%%%%%%%%%%%%%%%%%%%%%%%%%%%%%%%%%%%%%%%%%%%%%%%%%%%%%%%%

\section{Summary and Conclusions}
\label{Conclusion} 

We investigated the properties of boson stars with a generalized scalar potential by 
extending the power-law potential $V \propto \lambda \phi^n$ 
to an arbitrary value of the exponent $n$. We derive analytic expressions for the equation of state as input
to the TOV equations for static spheres of fluids.
We introduced two ways to stabilize the boson star configurations: by including a mass term for the bosons 
and by including a vacuum term in the scalar potential without a mass term.
For both cases we calculated the mass-radius curves, the speed of sound, as well as their maximal compactness. 
We found three different categories of the mass-radius curves for the EOS with a mass term. 
They differ in their behaviour in the limit of small masses. 
The radius $R$ for the case $n = 3$ goes to infinity when the mass $M$ goes to zero. 
For the classical $\phi^4$-potential, i.e. for $n = 4$, the radius goes to a constant value for small masses.
The mass-radius curves for larger values of $n$ resemble the mass-radius curves of self-bound stars 
as their mass vanishes when the radius goes to zero. 
Nevertheless, these compact star configurations are not self-bound because gravity is needed to stabilize them.
However, we find that the mass-radius curves for the EOS with a vacuum term turn out to be like those of self-bound stars. 
The mass vanishes for small radii but in this case the EOS has a nonvanishing energy density for a vanishing pressure
so that these compact star configurations are bound without gravity, i.e.\ they constitute self-bound star configurations.
For both types of EOSs studied we find that the speed of sound squared $c_s^2$ goes asymptotically to the limit of causality $c_s^2=1$. 
The case for the EOS with a vacuum constant and the one with a mass term case differ slightly in their absolute values 
for the same value of the power $n$ while the former one turns out to be always stiffer compared to the latter one.
Both cases show an increase of the speed of sound squared with the power $n$, so that higher powers in $n$ lead to a stiffer EOS. 
We calculated also the compactness for several values of the exponent $n$ in the range between $n = 3$ and $n = 100$. 
We demonstrate that the maximal compactness $C_\text{max}$ increases continuously with the power $n$. 
The highest increase occurs between $n = 3$ and $n = 10$. 
For higher values of $n$ the maximal compactness goes asymptotically to the limit of causality for both types of EOS, 
in line with the increase of the speed of sound squared $c_s^2$ with $n$ seen before.
The highest values of $C_\text{max}$ for a given power $n$ are reached with the EOS with a vacuum term. 
Even for low values of $n$ the compactness can be already close to the conformal limit, where $c_s^2 = 1/3$. 
Specifically, for the EOS with a vacuum term and $n = 4$ we reproduce the result from 
the equation of state of an ultrarelativistic ideal gas of quarks with a vacuum term, the MIT bag model, with a compactness of $C = 0.271$.
Furthermore, for the EOS with a mass term and the one with a vacuum term the maximal compactness 
can reach and even surpass a compactness of $C = 1/3$ marking the compactness needed to place the photon ring outside of the compact object. 
This makes boson stars described by a scalar potential with a power law with the power of $n \gtrsim 20$ black hole mimickers.
In summary, we establish that by extending the scalar potential $V = \lambda \phi^n$ to an arbitrary value of the exponent $n$ we change the properties, 
in particular the compactness, of the resulting compact configurations drastically. 
We point out that we have investigated properties like mass and radius of the boson star independently of 
the mass of the scalar field boson $m$ as well as the coupling strength $\lambda$, which enables a generic approach.
For a given dark matter model of self-interacting bosonic dark matter our results can be rescaled to physical units by the simple rescaling laws given 
in the derivation of the EOS. 
By choosing a suitable mass scale and interaction strength one can obtain boson stars with masses and radii comparable to 
(supermassive) black holes and neutron stars, for example. 
Our results could be used to study now boson star mergers with the generalized scalar potential. By choosing different values of the power of the scalar 
potential one is now able to study within the same setting boson star configurations with entirely different properties concerning the mass-radius relation, 
the speed of sound squared, and the maximal compactness up to black hole mimickers and to delineate 
the impact on the pattern of emitted gravitational waves and possible signals for their detection 
in present and future gravitational wave observatories.

%=============================================================================
\begin{acknowledgments}
 The authors acknowledge support by the Deutsche Forschungsgemeinschaft (DFG, German Research Foundation) through the CRC-TR 211 'Strong-interaction matter under extreme conditions'– project number 315477589 – TRR 211.
\end{acknowledgments}
%===============================================================================

\bibliographystyle{apsrev4-2} 
\bibliography{Paper_Refs.bib}

\end{document}